\documentclass[aps,prd,groupedaddress,twocolumn,floatfix,letterpaper]{revtex4-2}

\usepackage{anyfontsize}
\usepackage{bm}
\usepackage[hidelinks]{hyperref}
\usepackage{graphicx}
\usepackage{amsmath}
\usepackage{lipsum}
\usepackage{color}
\usepackage{lineno}

\definecolor{darkblue}{RGB}{0,0,196}
\definecolor{darkred}{RGB}{196,0,0}
\definecolor{darkgreen}{RGB}{0,136,0}

\begin{document}

\title{Transverse momentum dependent feed-down fractions for bottomonium production}
\author{Jacob Boyd}
\author{Michael Strickland}
\author{Sabin Thapa}
\affiliation{Kent State University, Department of Physics, Kent, OH 44242 USA}

\begin{abstract}
We extract transverse momentum dependent feed-down fractions for bottomonium production using a data-driven approach.  We use data published by the ATLAS, CMS, and LHCb Collaborations for $\sqrt{s} =7 $ TeV proton-proton collisions.  Based on this collected data, we produce fits to the differential cross sections for the production of both S- and P-wave bottomonium states.  Combining these fits with branching ratios for excited state decays from the Particle Data Group, we compute the feed-down fractions for both the $\Upsilon(1S)$ and $\Upsilon(2S)$ as a function of transverse momentum.  Our results indicate a strong dependence on transverse momentum, which is consistent with prior extractions of the feed-down fractions.  When evaluated at the average momentum of the states, we find that approximately 75\% of $\Upsilon(1S)$ and $\Upsilon(2S)$ states are produced directly.  Our results for the transverse momentum dependent feed-down fractions are provided in tabulated form so that they can be used by other research groups. 
\end{abstract}


\maketitle

\section{Introduction}

The suppression of bottomonium production in heavy-ion collisions is a key signature for the production of a deconfined quark-gluon plasma in ultra-relativistic heavy-ion collisions \cite{Mocsy:2013syh,Andronic:2015wma,Strickland:2021boy,STAR:2013kwk,PHENIX:2014tbe,STAR:2016pof,CMS:2017ycw,Sirunyan:2018nsz,ALICE:2019pox,Acharya:2020kls,Lee:2021vlb,CMS:2020efs,CMS:2022rna}.  There are now various approaches to compute this suppression, which include manifestly quantum mechanical approaches \cite{Laine:2006ns,Beraudo:2007ky,Brambilla:2008cx,Escobedo:2008sy,Brambilla:2010vq,Akamatsu:2011se,Strickland:2011mw,Strickland:2011aa,Akamatsu:2014qsa,Blaizot:2015hya,Krouppa:2015yoa,Katz:2015qja,Brambilla:2016wgg,Krouppa:2016jcl,Blaizot:2017ypk,Brambilla:2017zei,Krouppa:2017jlg,Yao:2018nmy,Brambilla:2019tpt,Rothkopf:2019ipj,Islam:2020gdv,Islam:2020bnp,Brambilla:2020qwo,Akamatsu:2020ypb,Yao:2020xzw,Yao:2020eqy,Brambilla:2021wkt,Omar:2021kra,Blaizot:2021xqa,Yao:2021lus,Brambilla:2022ynh,Alalawi:2022gul,Brambilla:2023hkw,Strickland:2023nfm} and others based on kinetic transport equations \cite{Grandchamp:2005yw,Rapp:2008tf,Du:2017hss,Yao:2018nmy,Du:2019tjf,Hatwar:2020esf,Yao:2020xzw,Yao:2020eqy}.  One common feature of phenomenological approaches is that, in order to compare to experimental data, they must include a late-time feed-down stage in order to account for the decays of excited bottomonium states to lower-lying states.  Typically, this is done using momentum-independent feed-down fractions, see e.g. Sec. 6.4 of Ref.~\cite{Brambilla:2020qwo} or Sec.~4.3 of Ref.~\cite{Yao:2020xzw}.

In this paper we make an analysis of the world's collected data on bottomonium production in pp collisions at $\sqrt{s}$ = 7 TeV and make a data-driven extraction of the feed-down fractions for the $\Upsilon(1S)$ and $\Upsilon(2S)$.  We make fits to the S-wave scattering cross sections as measured by the ATLAS \cite{ATLAS:2012lmu} and CMS \cite{CMS:2013qur,CMS:2015xqv} Collaborations and then use these together with data on the P-wave production cross sections provided by the LHCb Collaboration \cite{LHCb:2014ngh}.  Based on these fits and branching fractions for various excited state decays available in Particle Data Group  listings, we compute the transverse momentum dependent feed-down fractions.  

We make two independent extractions of the transverse momentum, $p_T$, dependent feed-down fractions based on data from ATLAS and CMS Collaborations, allowing a consistency check of the final results.  We find that both the ATLAS- and CMS-based feed-down fractions have a strong dependence on transverse momentum, which is numerically consistent with results presented previously by H. W\"{o}hri~\cite{Wohri:2015hq}.  We find that the feed-down fractions extracted from the ATLAS and CMS Collaborations are largely compatible, however, some residual dependence on which set of S-wave data is used exists.  We provide tabulated results for the feed-down fractions based on both the ATLAS and CMS data so that other research groups can gauge the sensitivity to which set of experimental data is used in the extraction.  

We will demonstrate that, independent of whether the ATLAS or CMS experimental data sets are used, when evaluated at the average transverse momentum of the 1S and 2S states, the percentage of both the 1S and 2S states produced directly is on the order of 75\%.  This rules out the often speculated possibility that the suppression of ground state $\Upsilon$ production in nucleus-nucleus collisions could be solely due to excited state suppression.  This speculation was based on an outdated understanding of bottomonium feed-down fractions.  Using the results contained herein, a unified picture of bottomonium feed down can be implemented by all research groups studying this topic, reducing this uncertainty in the analysis of bottomonium suppression in nucleus-nucleus collisions.  
In addition, our results for the feed-down fractions may be used in the computation of other observables of interest in high-energy physics.  Finally, we note that, although we consider $\sqrt{s} = 7$ TeV collisions herein, the linear scaling of bottomonium production cross sections with the collision energy at LHC energies means that our results can be applied at lower energies, e.g., $\sqrt{s} = 5.02$ TeV, which are relevant for heavy-ion collisions.

The structure of our paper is as follows.  In Sec.~\ref{sec:swavefits}, we present the world's collected data for S-wave bottomonium production in $\sqrt{s}$ = 7 TeV pp collisions and make fits to this data in order to construct smooth functions that can be used at all $p_T$.  In Sec.~\ref{sec:pwavefits}, we review LHCb measurements of the $\chi_{b}(mP) \rightarrow \Upsilon(nS)\gamma$ process and discuss how these measurements, together with the S-wave production measurements of ATLAS and CMS, can be used to obtain the differential cross sections for $\chi_{b}(mP)$ production.  In Sec.~\ref{sec:xsecs}, we present extractions of the $p_T$-integrated cross sections and average $p_T$ for $\Upsilon(nS)$ and $\chi_b(mP)$ states.  In Sec.~\ref{sec:feeddownfracs}, we present our final results for the $p_T$-dependent feed-down fractions obtained using both ATLAS and CMS S-wave production data.  Finally, in Sec.~\ref{sec:conclusions}, we present our conclusions and an outlook for the future.  We list a particularly unwieldy expression for a matrix composed of all bottomonium branching fractions in App.~\ref{sec:feeddownmatrix}.

\section{Fits to ATLAS and CMS S-wave bottomonium cross sections}
\label{sec:swavefits}

\begin{figure}[t]
\vspace{-1mm}
\begin{center}
\includegraphics[width=0.95\linewidth]{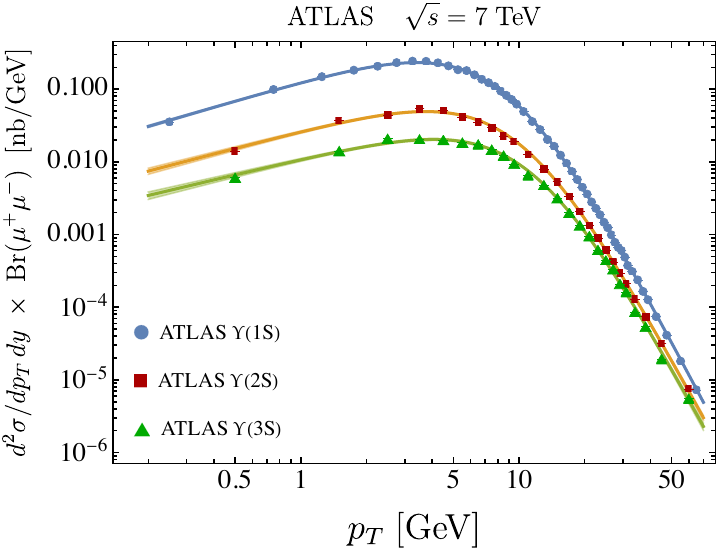} \\[1em]
\includegraphics[width=0.95\linewidth]{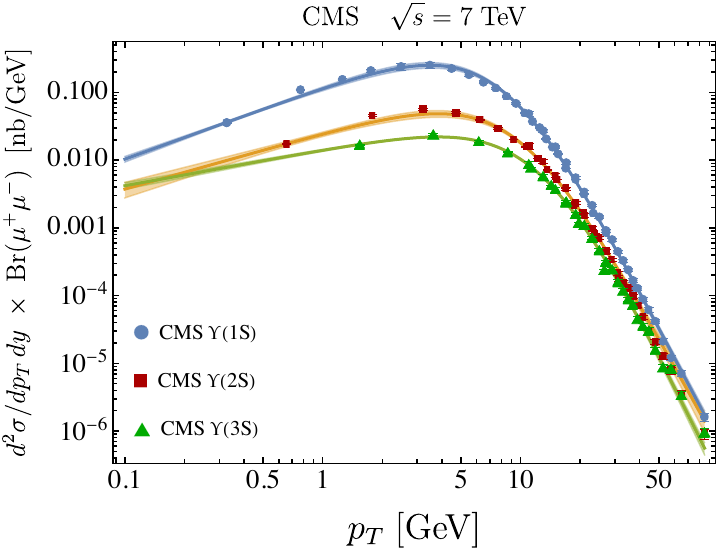}
\end{center}
\vspace{-3mm}
\caption{Differential cross sections for $\Upsilon(1S)$, $\Upsilon(2S)$, and $\Upsilon(3S)$ production in $pp$ collisions at $\sqrt{s}=$ 7 TeV as a function of transverse momentum $p_{T}$.  Results for each state are multiplied by their corresponding branching fractions to dimuons.  The top panel shows the fit obtained using data from ATLAS \cite{ATLAS:2012lmu} and the bottom panel shows the fit obtained using data from CMS \cite{CMS:2013qur,CMS:2015xqv}.
} 
\label{fig:1s2s3sfit}
\end{figure}

In Fig.~\ref{fig:1s2s3sfit} we plot ATLAS \cite{ATLAS:2012lmu} and CMS \cite{CMS:2013qur,CMS:2015xqv} data for the differential cross sections for $\Upsilon(1S)$, $\Upsilon(2S)$, and $\Upsilon(3S)$ production times their respective dimuon branching fractions of $0.0248\pm 0.0004$, $0.0193\pm 0.0017$, and $0.0218\pm 0.0021$~\cite{pdg}.  The data are plotted on a log-log scale in order to call the readers attention to the fact that the experimental results indicate that there exists two different power law scalings, one appropriate for low momentum and one for high momentum.  In order to capture this feature of the data, we make a fit to each of the S-wave data sets using a functional form 
\begin{equation}
f(p_T, a, b, c, d) = \frac{a \, p_{T}^{b}}{\left(p_{T}^{2}+c^{2}\right)^{d}} \, ,
\end{equation}
where $a$, $b$, $c$, and $d$ are fit parameters.  For transverse momentum $p_T \ll c$, the exponent of $p_T$ is given by $b$ and for transverse momentum $p_T \gg c$, it is given by $b-2d$.  Using this form, we made non-linear fits to the experimental data from both collaborations, with the results plotted in Fig.~\ref{fig:1s2s3sfit} as solid lines surrounded by shaded bands.  The shaded bands correspond to the 1$\sigma$ confidence intervals determined self-consistently from the fit.  

The 1$\sigma$ bands shown in Fig.~\ref{fig:1s2s3sfit} were determined by computing the covariance matrix $M = {\rm var}(P)$, where $P$ is a vector containing the fit parameters used. We then left- and right-contracted the covariance matrix with a vector $V$ containing the derivatives of the fit function with respect to all fit parameters, which were evaluated at the best fit parameters.  The upper and lower boundaries of the bands are determined by taking the best fit function plus or minus the square root of this contraction, i.e.,  $f_{\rm bands}^\pm(p_T) = f_{\rm best\;fit}(p_T) \pm \sqrt{V^T(p_T)\cdot M \cdot V(p_T)}$.

As can be seen from Fig.~\ref{fig:1s2s3sfit}, this fit form provides a good description of both the low- and high-$p_T$ forms of the observed S-wave differential cross sections.  We list the extracted best-fit parameters in Tab.~\ref{tab:swavefit} along with the associated $\chi^2$ per degree of freedom.  We note that, when compared to the fits to the ATLAS Collaboration data, the larger $\chi^2$ per degree of freedom for the CMS 1S and 2S fits are due to tension between the two reported CMS data sets and the presence of some significant outliers in their reported data \cite{CMS:2013qur,CMS:2015xqv}.  In addition to this, one observes that the CMS data has much smaller reported uncertainties than reported by the ATLAS Collaboration.  Despite this, as can be seen in Fig.~\ref{fig:1s2s3sfit} the fits to both the ATLAS and CMS data describe the data quite well given the simple form of the fit function used.

\begin{table}[t!]
\begin{center}
\begin{tabular}{|c|c|c|c|c|c|}
 \hline
 \multicolumn{6}{|c|}{\bf ATLAS} \\
 \hline
\;\textbf{State}\; & \;\textbf{a}$\bm{/10^5}$\; & \;\;\;\; \textbf{b} \;\;\;\; & \;\;\;\; \textbf{c} \;\;\;\; & \;\;\;\; \textbf{d} \;\;\;\; & \;\; $\chi^2/{\rm dof}$ \;\; \\[0.25em] \hline
 $\Upsilon(1S)$ & 2.111 & 0.8806 & 8.673 & 3.316 & 0.290 \\[0.25em] \hline
 $\Upsilon(2S)$ & 0.5263 & 0.7875 & 9.906 & 3.162 & 0.980 \\[0.25em] \hline
 $\Upsilon(3S)$ & 0.4350 & 0.7158 & 11.300 & 3.134 & 1.18 \\[0.25em] \hline
\end{tabular}
\\[1.5em]
\begin{tabular}{|c|c|c|c|c|c|}
 \hline
 \multicolumn{6}{|c|}{\bf CMS} \\
 \hline
\;\textbf{State}\; & \;\textbf{a}$\bm{/10^5}$\; & \;\;\;\; \textbf{b} \;\;\;\; & \;\;\;\; \textbf{c} \;\;\;\; & \;\;\;\; \textbf{d} \;\;\;\; & \;\; $\chi^2/{\rm dof}$ \;\; \\[0.25em] \hline
 $\Upsilon(1S)$ & 1.429 & 1.061 & 8.042 & 3.357 & 29.1 \\[0.25em] \hline
 $\Upsilon(2S)$ & 0.3267 & 0.8247 & 9.743 & 3.093 & 81.6  \\[0.25em] \hline
 $\Upsilon(3S)$ & 3.094 & 0.5310 & 12.94 & 3.299 & 1.13 \\[0.25em] \hline
\end{tabular}
\vspace{1mm}
\caption{Best fit parameters for the S-wave differential cross sections shown in Fig.~\ref{fig:1s2s3sfit}.}
\label{tab:swavefit}
\end{center}
\end{table}

\begin{figure}[t]
\vspace{-1mm}
\begin{center}
\includegraphics[width= 0.95\linewidth]{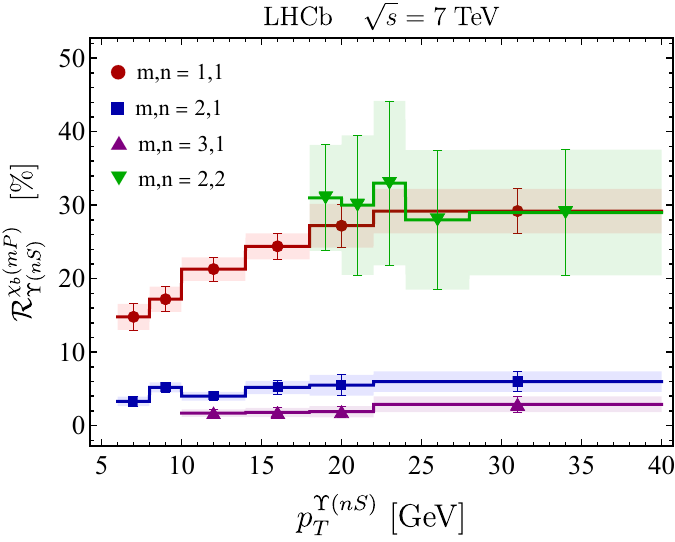}
\end{center}
\vspace{-3mm}
\caption{Relative production of P-wave states to S-wave states, $R^{\chi_{b} (mP)}_{\Upsilon (nS)}$, as a function of $p_{T}^{\Upsilon}$ for pp collisions at $\sqrt{s}$ =  7 TeV.  The results shown above summarize the ratios reported by the LHCb Collaboration in Ref.~\cite{LHCb:2014ngh}.
} 
\label{fig:ratioplot}
\end{figure}

\begin{figure*}[t]
\begin{center}
\includegraphics[width=0.45\linewidth]{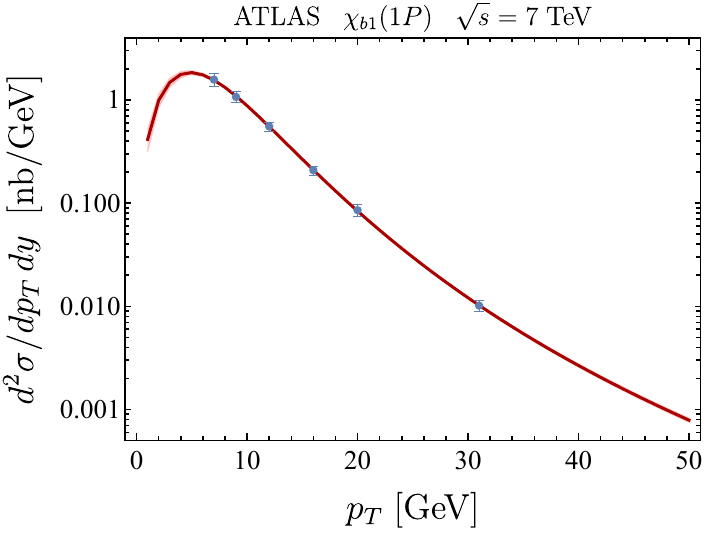} 
\;\;\;\;\;
\includegraphics[width=0.45\linewidth]{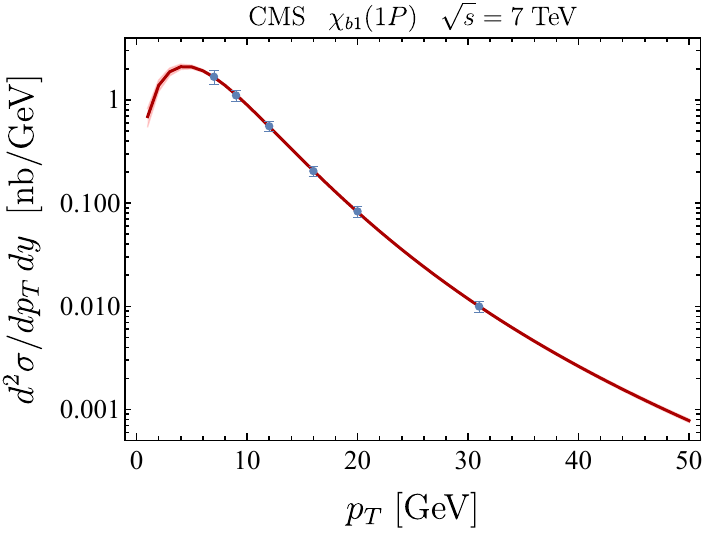}  
\\[1em]
\includegraphics[width=0.45\linewidth]{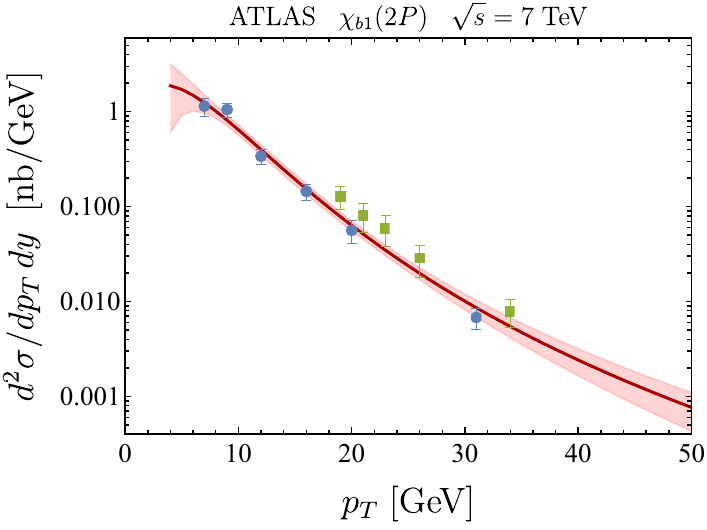}
\;\;\;\;\;
\includegraphics[width=0.45\linewidth]{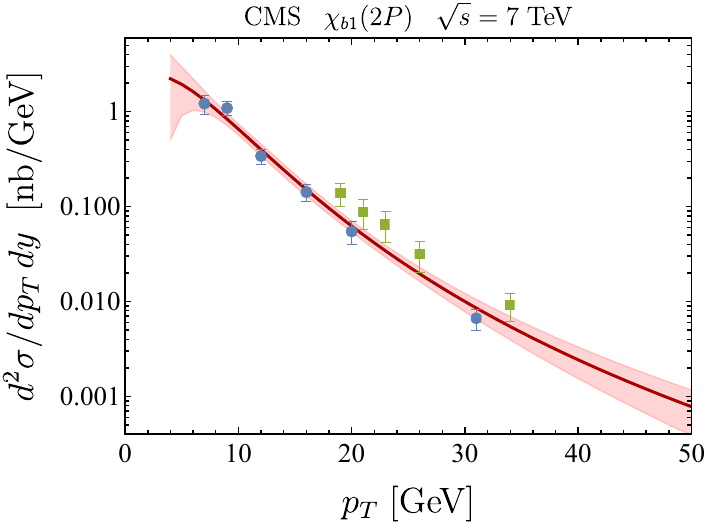}
\end{center}
\vspace{-4mm}
\caption{Differential cross sections for the $\chi_{b1}(1P)$ (top row) and $\chi_{b1}(2P)$ (bottom row) in $\sqrt{s} = $ 7 TeV pp collisions as a function of $p_{T}$.  The left column shows results obtained using the ratios measured by the LHCb Collaboration in Ref.~\cite{LHCb:2014ngh} together with S-wave production data from ATLAS \cite{ATLAS:2012lmu}.  The right column shows results obtained using the ratios measured by the LHCb Collaboration in Ref.~\cite{LHCb:2014ngh} together with S-wave production data from CMS \cite{CMS:2013qur,CMS:2015xqv}.  In the bottom row, data inferred using $R_{\Upsilon(1S)}^{\chi_{b}(2P)}$ are indicated by blue filled circles, while data inferred using $R_{\Upsilon(2S)}^{\chi_{b}(2P)}$ are indicated as green filled squares.}
\label{fig:1p2p}
\end{figure*}

\section{LHCb P-wave to S-wave ratio measurements}
\label{sec:pwavefits}

In order to compute the feed-down fractions, we need to make fits to all available data for P-wave bottomonium production as well.
For this purpose, we can make use of LHCb Collaboration $\sqrt{s}$ = 7 TeV pp collision data for the ratio of P-wave to S-wave bottomonium collected in the rapidity range $2.0 < y < 4.5$ \cite{LHCb:2014ngh}.  In Ref.~\cite{LHCb:2014ngh} the collaboration reported the fraction of $\Upsilon(nS)$ mesons produced via $\chi_{b} \rightarrow \Upsilon (nS)\gamma$ decays as a function of the $\Upsilon(nS)$ transverse momentum.  In Fig.~\ref{fig:ratioplot}, we summarize the results obtained by the LHCb Collaboration.  Using this information together with our prior fits to ATLAS and CMS data, one can obtain estimates for the differential cross sections for P-wave bottomonium production. 

The ratios reported by the LHCb Collaboration are given by~\cite{LHCb:2014ngh}
\begin{eqnarray}
    R^{\chi_{b} (mP)}_{\Upsilon (nS)} &\equiv& \frac{\sigma(pp \rightarrow \chi_{b1}(mP)X)}{\sigma(pp \rightarrow \Upsilon(nS)X)} \times B_{1} \nonumber \\ && \hspace{1cm} + \frac{\sigma(pp \rightarrow \chi_{b2}(mP)X)}{\sigma(pp \rightarrow \Upsilon(nS)X)} \times B_{2} \, ,
    \label{eq:rform}
\end{eqnarray} 
where, $B_{1(2)}$ are the branching ratios for $\chi_{b1(2)} \rightarrow \Upsilon(nS) \gamma$.  For the $\chi_{b}(1P) \rightarrow \Upsilon(1S)\gamma$, we used the values $B_1 = 0.352 \pm 0.020$ and $B_2 = 0.18 \pm 0.01$.  For $\chi_{b}(2P) \rightarrow \Upsilon(1S)\gamma$ , we use the values $B_1 = 0.099 \pm 0.010$ and $B_2 = 0.066 \pm 0.008$.  Finally, for the $\chi_{b}(2P) \rightarrow \Upsilon(2S)\gamma$, we used the values $B_1 = 0.181 \pm 0.019$ and $B_2 = 0.089 \pm 0.012$.  In all cases, the values come from the Particle Data Group (PDG) listings \cite{pdg}.  Contributions from $\chi_{b0}(mP) \rightarrow \Upsilon(nS) \gamma$ decays are neglected in Eq.~\eqref{eq:rform} due to their small branching fraction \cite{pdg,LHCb:2014ngh} \footnote{For $\chi_{b0}(1P) \rightarrow \Upsilon(1S)\gamma$, $\chi_{b0}(2P) \rightarrow \Upsilon(1S)\gamma$, and $\chi_{b0}(2P) \rightarrow \Upsilon(2S)\gamma$ the PDG branching fractions are $0.0194 \pm 0.0027$, $(3.8 \pm 1.7) \times 10^{-3}$, and $0.0138 \pm 0.003$, respectively.}.

Introducing the ratio 
\begin{equation}
r_{21}(mP) \equiv \frac{\sigma(pp \rightarrow \chi_{b2}(mP)X)}{\sigma(pp \rightarrow \chi_{b1}(mP)X)} \; ,
\label{eq:r21}
\end{equation} 
one can obtain the pp cross section for $\chi_{b1}(mP)$ production using the following relation
\begin{equation}
\sigma(pp \rightarrow \chi_{b1}(mP)X) =   \frac {R_{\Upsilon(nS)}^{\chi_{b}(mP)} \times \sigma(pp \rightarrow \Upsilon(nS)X)}{B_{1} + r_{21} \times B_{2}} \, .
\label{eq:sigmap}
\end{equation}

The value of $r_{21}(1P)$ is constrained experimentally to be $r_{21}(1P) = 0.85 \pm 0.15$ \cite{Khachatryan:2014ofa}.  Once $\sigma(pp \rightarrow \chi_{b1}(mP)X)$ is determined, we can use Eq.~\eqref{eq:r21} to determine the $pp \rightarrow \chi_{b2}(mP)X$ cross sections \footnote{Here we assume that $r_{21}(2P) = r_{21}(1P)$.}.  Finally, due to the fact that there exists no experimental data for the production of $\chi_{b0}(mP)$ at LHC energies, to determine the $pp \rightarrow \chi_{b0}(mP)X$ cross sections, we make use of expectations from non-relativistic QCD to set the $\chi_{b0}(mP)$ production cross section to 1/4 of the average of the $\chi_{b1}(mP)$ and $\chi_{b2}(mP)$ cross sections.  This estimate is based on the expectation that $\chi_{b}(mP)$ production comes mainly from the color-octet channel and is proportional to the spin multiplicity $2J+1$ of the state~\cite{Baier:1983va,Brambilla:2022ayc,HeeSok}.

\begin{table}[t!]
\begin{center}
\begin{tabular}{|c|c|c|c|c|}
 \hline
 \multicolumn{5}{|c|}{\bf ATLAS + LHCb} \\
 \hline
\;\;\; \textbf{State} \;\;\; & \;\textbf{a}$\bm{/10^5}$\; & \;\;\;\; \textbf{b} \;\;\;\; & \;\;\;\; \textbf{c} \;\;\;\; & \;\; $\chi^2/{\rm dof}$ \;\;\\[0.25em] \hline
 $\chi_{b1}(1P)$ & 65.45 & 1.413 & 3.608 & 0.0213 \\[0.25em] \hline
 $\chi_{b1}(2P)$ & 15.67 & 0.5621 & 3.005 & 1.32 \\[0.25em] \hline
\end{tabular}
\\[1em]
\begin{tabular}{|c|c|c|c|c|}
 \hline
 \multicolumn{5}{|c|}{\bf CMS + LHCb} \\
 \hline
\;\;\; \textbf{State} \;\;\; & \;\textbf{a}$\bm{/10^5}$\; & \;\;\;\; \textbf{b} \;\;\;\; & \;\;\;\; \textbf{c} \;\;\;\; & \;\; $\chi^2/{\rm dof}$ \;\;\\[0.25em] \hline
 $\;\chi_{b1}(1P)\;$ & 54.53 & 1.155 & 3.458 & 0.0152 \\  \hline
 $\chi_{b1}(2P)$ & 12.00 & 0.2962 & 2.837 & 1.64 \\ \hline
\end{tabular}
\vspace{1mm}
\caption{Best fit parameters corresponding to the differential cross sections fits for $\chi_{b1}$ production shown Fig.~\ref{fig:1p2p}.}
\label{tab:pwavefit}
\end{center}
\end{table}

Using Eq.~\eqref{eq:sigmap} and S-wave data from either the ATLAS or the CMS Collaborations, one can compute the $\chi_{b1}(mP)$ differential cross sections.  We note that, as can be seen in Fig.~\ref{fig:ratioplot},  $R_{\Upsilon(nS)}^{\chi_{b}(mP)}$ depends strongly on transverse momentum below $p_T \sim 20$ GeV and hence the differential cross sections for $\chi_{b}$ production will depend on $p_{T}$ in a different manner than the S-wave states at low $p_T$.  In Fig.~\ref{fig:1p2p} we plot the $p_T$-differential cross sections obtained for the $\chi_{b1}(1P)$ (top row) and $\chi_{b1}(2P)$ (bottom row) in $\sqrt{s} = $ 7 TeV pp collisions.  The left column shows results obtained using the ratios measured by the LHCb Collaboration in Ref.~\cite{LHCb:2014ngh} together with S-wave production data from ATLAS \cite{ATLAS:2012lmu}.  The right column shows results obtained using the ratios measured by the LHCb Collaboration in Ref.~\cite{LHCb:2014ngh} together with S-wave production data from CMS \cite{CMS:2013qur,CMS:2015xqv}.

In order to fit the $\chi_{b1}(1P)$ and $\chi_{b1}(2P)$ production data shown in Fig.~\ref{fig:1p2p} we use the following fit function
\begin{equation}
g(p_T, a, b, c) = \frac{a \times p_{T}^b}{(p_{T}^2 + \tilde{m}^2)^c} \; ,
\label{eq:pwavefitform}
\end{equation}
where we take $\tilde{m} = 9.983$ GeV for the $\chi_{b1}(1P)$ and $\tilde{m} = 10.255$ GeV for the $\chi_{b1}(2P)$, which are the central values of their masses from the PDG, respectively.  We use $a$, $b$, and $c$ as fit parameters.  We have reduced the number of fit parameters in this case due the fact that there are many fewer data points available than for the S-wave states.  
Our results are shown as red solid lines with shaded red bands indicating the $1\sigma$ confidence interval.
In the case of the $\chi_{b1}(1P)$, the precision of the experimental data allows for a rather constrained fit as can be seen in the top row of Fig.~\ref{fig:1p2p}.  

\begin{figure}[t]
\begin{center}
\includegraphics[width= 0.95\linewidth]{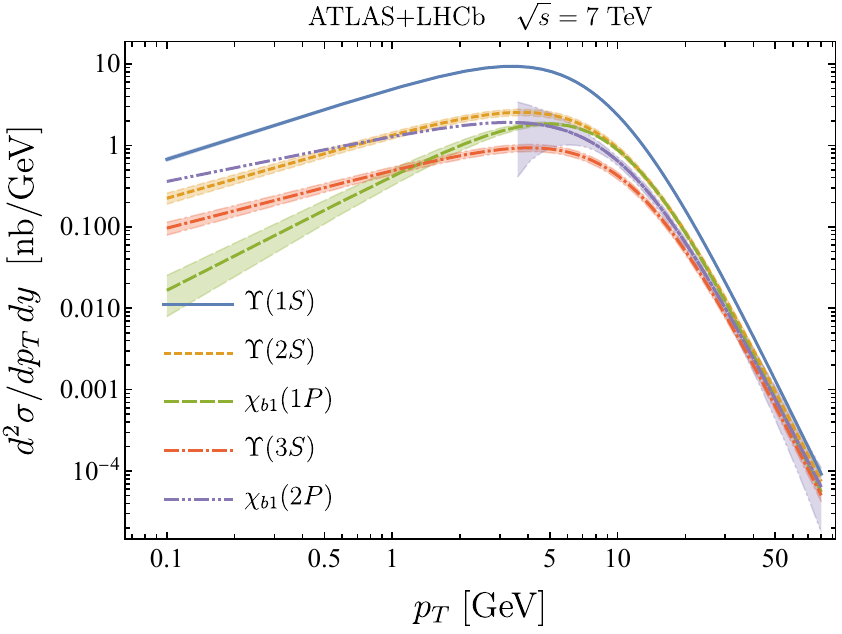}\\[1em]
\includegraphics[width= 0.95\linewidth]{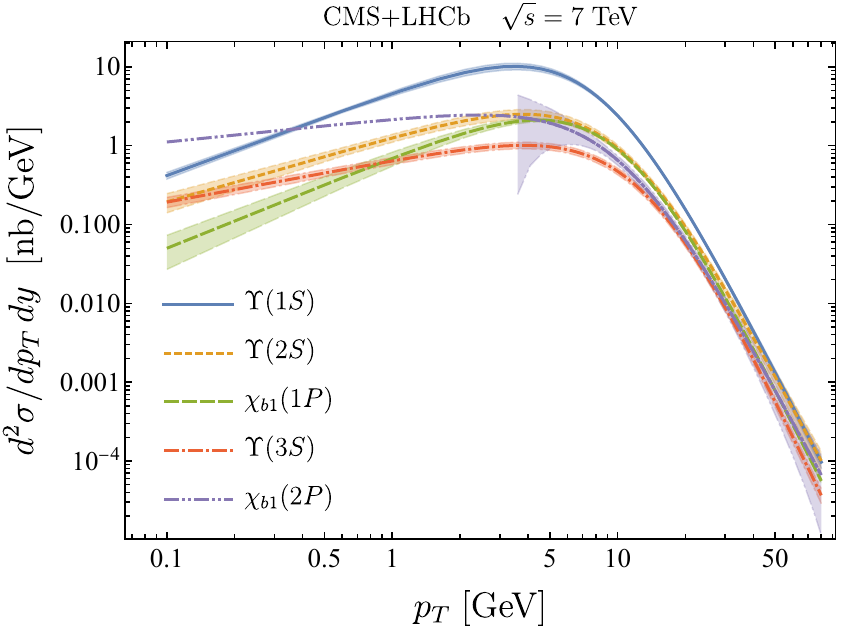}
\end{center}
\vspace{-3.5mm}
\caption{Fitted differential cross sections for $\Upsilon(1S)$, $\Upsilon(2S)$, $\chi_{b1}(1P)$, $\Upsilon(3S)$, and $\chi_{b1}(2P)$ states as a function of $p_T$.  The $\Upsilon(1S)$, $\Upsilon(2S)$, and $\Upsilon(3S)$ cross sections are extracted from ATLAS data \cite{ATLAS:2012lmu} and CMS data and the $\chi_{b1}(1P)$ and $\chi_{b1}(2P)$ cross sections are extracted using these cross sections combined with LHCb data~\cite{LHCb:2014ngh}.}
\label{fig:finalxsecs}
\end{figure}

In the case of the $\chi_{b1}(2P)$ production data, one can infer these using either the $\Upsilon(1S)$ or $\Upsilon(2S)$ cross sections and these two possibilities are indicated by blue filled circles or green filled squares in Fig.~\ref{fig:1p2p}, respectively.  Due to the differences in the two results, the fit in this case is less constrained.  We do not show results below \mbox{$p_T =$ 4 GeV} in Fig.~\ref{fig:1p2p} due the fact that the lower limit on the confidence interval becomes negative, precluding it from being displayed on a logarithmic scale.  We note, however, that internally we use the resulting fit below this scale in our analysis despite the fact that it cannot be plotted on a logarithmic scale.  The resulting best-fit values for the parameters $a$, $b$, and $c$ for both the $\chi_{b1}(1P)$ and $\chi_{b1}(2P)$ are listed in Tab.~\ref{tab:pwavefit} along with the associated $\chi^2$ per degree of freedom.  We note that small reported $\chi^2$ per degree of freedom for the $\chi_{b1}(1P)$ states results from the fact that there are quite few LHCb data points available for $R^{\chi_{b} (mP)}_{\Upsilon (nS)}$ and these data points are tightly constrained.

\section{Integrated cross sections and average transverse momentum}
\label{sec:xsecs}

Based on the fits obtained in the previous two Sections, we have a description of the most relevant states necessary for computing the momentum-dependent feed-down fractions for the $\Upsilon(1S)$ and $\Upsilon(2S)$ states.  As a precursor to this, in Fig.~\ref{fig:finalxsecs} we present the fitted differential cross sections for production of the $\Upsilon(1S)$, $\Upsilon(2S)$, $\chi_{b1}(1P)$, $\Upsilon(3S)$, and $\chi_{b1}(2P)$ states in $\sqrt{s} = 7$ TeV pp collisions as a function of $p_T$.  In Fig.~\ref{fig:finalxsecs}, the top panel shows results obtained using ATLAS and LHCb data and the bottom panel shows results obtained using CMS and LHCb data.  As before, the shaded bands correspond to $1 \sigma$ confidence intervals for the fits.  

Note that, for the $\chi_{b1}(2P)$, we terminate the display of the confidence interval for the $\chi_{b1}(2P)$ at $p_T = $ 4 GeV due to the fact that the lower limit becomes negative and cannot be plotted on a logarithmic scale; however, we continue the central line in this figure, which corresponds to the best fit parameters listed in Tab.~\ref{tab:pwavefit}.  In practice, when computing the $p_T$-integrated cross sections, average $p_T$, and feed-down fractions, we use the full $1\sigma$ bands even if the confidence interval includes negative values \footnote{We have checked that reducing the confidence interval such that the lower limit of the interval remains positive does not significantly affect our final results.}.

\begin{table}[t!]
\begin{center}
\begin{tabular}{|c|c|c|} 
 \hline
 \multicolumn{3}{|c|}{\bf ATLAS + LHCb} \\
 \hline
 {\bf State} & \; $\bm{d\sigma/dy}$ \bf [nb] \; & \; $\bm{\langle p_T \rangle}$ \bf [GeV] \; \\[0.25em] \hline
 $\Upsilon(1S)$ & $69.069\pm 0.035$	&	$5.771\pm 0.010$  \\[0.25em] \hline
 $\Upsilon(2S)$ & $21.79\pm 0.05$	&	$6.60\pm 0.05$ \\[0.25em] \hline
 $\; \chi_{b0}(1P) \; $ & $3.81\pm 0.31$	&	$7.597\pm 0.031$ \\[0.25em] \hline
 $\chi_{b1}(1P)$ & $16.466\pm 0.025$	&	$7.597\pm 0.031$ \\[0.25em] \hline
 $\chi_{b2}(1P)$ & $14.0\pm 2.5$	&	$7.597\pm 0.031$  \\[0.25em] \hline
 $\Upsilon(3S)$ & $8.968\pm 0.022$	&	$7.32\pm 0.06$ \\[0.25em] \hline
 $\chi_{b0}(2P)$ & $3.80\pm 0.32$	&	$6.4\pm 0.4$ \\[0.25em] \hline
 $\chi_{b1}(2P)$ & $16.4\pm 0.4$	&	$6.4\pm 0.4$ \\[0.25em] \hline
 $\chi_{b2}(2P)$ & $14.0\pm 2.5$	&	$6.4\pm 0.4$ \\[0.25em] \hline
\end{tabular}
\\[1.5em]
\begin{tabular}{|c|c|c|} 
 \hline
 \multicolumn{3}{|c|}{\bf CMS + LHCb} \\
 \hline
 {\bf State} & \; $\bm{d\sigma/dy}$ \bf [nb] \; & \; $\bm{\langle p_T \rangle}$ \bf [GeV] \; \\[0.25em] \hline
 $\Upsilon(1S)$ & $72.34\pm 0.17$	&	$5.75\pm 0.04$  \\[0.25em] \hline
 $\Upsilon(2S)$ & $21.79\pm 0.07$	&	$6.76\pm 0.07$ \\[0.25em] \hline
 $\; \chi_{b0}(1P) \; $ & $4.28\pm 0.35$	&	$7.142\pm 0.030$ \\[0.25em] \hline
 $\chi_{b1}(1P)$ & $18.492\pm 0.029$	&	$7.142\pm 0.030$ \\[0.25em] \hline
 $\chi_{b2}(1P)$ & $15.7\pm 2.8$	&	$7.142\pm 0.030$  \\[0.25em] \hline
 $\Upsilon(3S)$ & $10.065\pm 0.023$	&	$7.23\pm 0.05$ \\[0.25em] \hline
 $\chi_{b0}(2P)$ & $4.6\pm 0.4$	&	$5.7\pm 0.6$ \\[0.25em] \hline
 $\chi_{b1}(2P)$ & $19.7\pm 0.8$	&	$5.7\pm 0.6$ \\[0.25em] \hline
 $\chi_{b2}(2P)$ & $16.8\pm 3.0$	&	$5.7\pm 0.6$ \\[0.25em] \hline
\end{tabular}
\end{center}
\caption{Transverse momentum integrated cross sections and average $p_T$ for bottomonium states in $\sqrt{s} = 7$ TeV pp collisions.  The top set of results is based on fits to ATLAS data \cite{ATLAS:2012lmu} for the S-wave states and LHCb data for the P-wave to S-wave ratios \cite{LHCb:2014ngh}.  The bottom set of results is based on fits to CMS data for S-wave ratios \cite{CMS:2013qur,CMS:2015xqv} and LHCb data for the P-wave to S-wave ratios.}
\label{tab:sevenxsec}
\end{table}

\begin{figure*}[t]
\begin{center}
\includegraphics[width=0.45\linewidth]{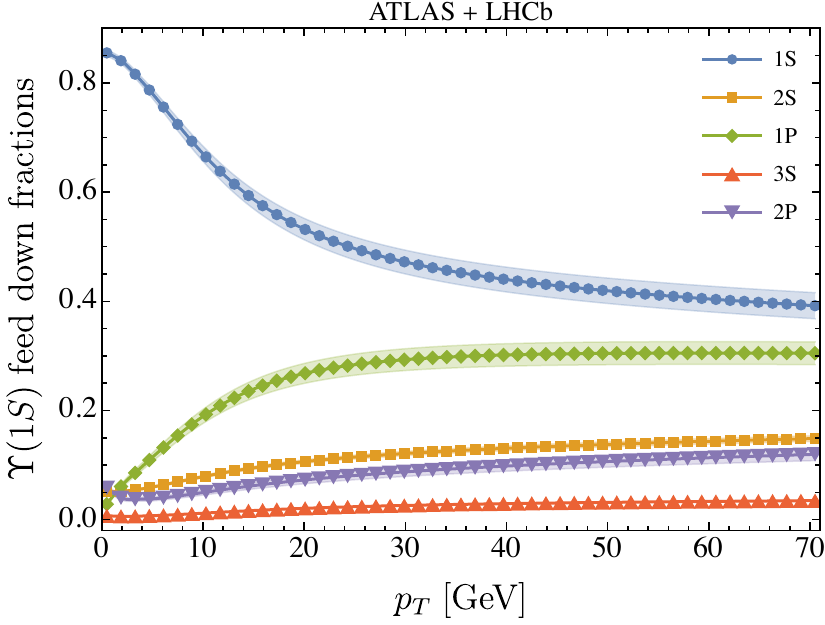} 
\;\;\;\;\;
\includegraphics[width=0.45\linewidth]{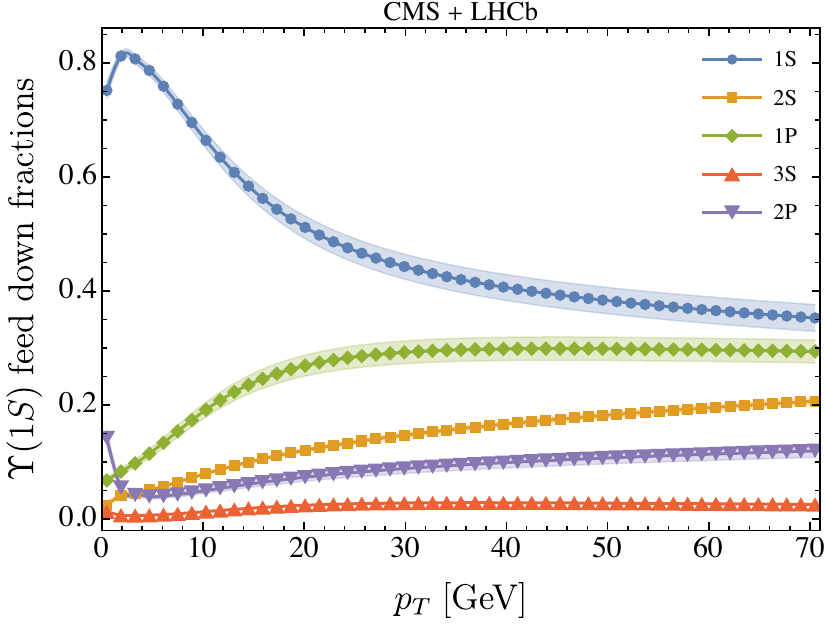}\\[1em]
\includegraphics[width=0.45\linewidth]{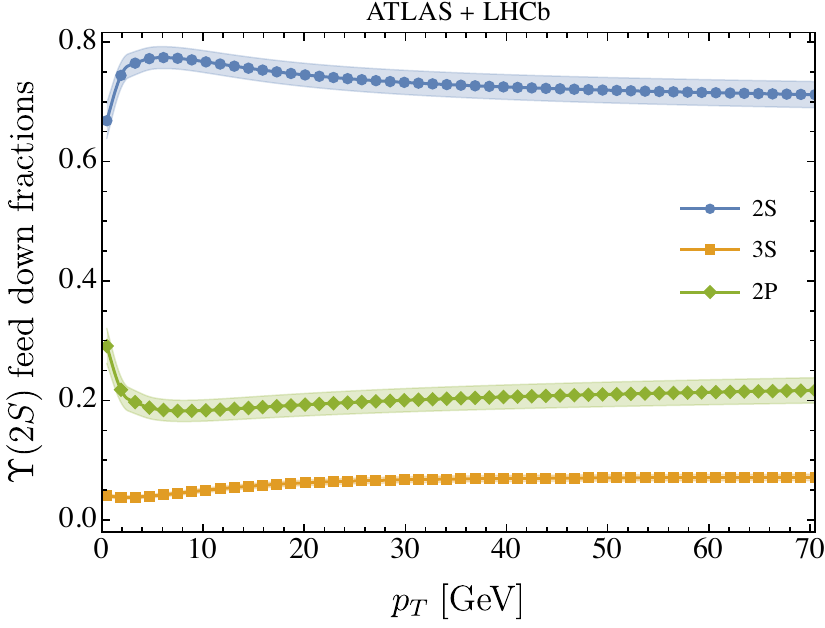}
\;\;\;\;\;
\includegraphics[width=0.45\linewidth]{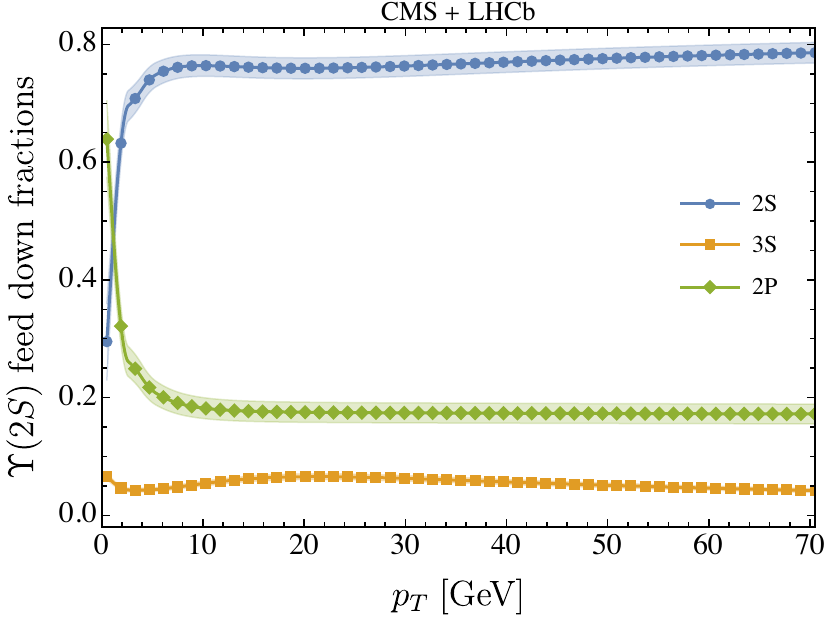}
\end{center}
\vspace{-4mm}
\caption{$\Upsilon(1S)$ (top row) and $\Upsilon(2S)$ (bottom row) feed-down fractions as a function of $p_T$ extracted from combined data from the ATLAS and LHCb Collaborations (left column) and data from the ATLAS and CMS Collaborations (right column)} 
\label{fig:feeddown}
\end{figure*}

Based on these fits one can obtain estimates for the $p_T$-integrated cross sections and the average $p_T$ for each state considered.  We present these results in Tab.~\ref{tab:sevenxsec}.
In the top set of results presented in Tab.~\ref{tab:sevenxsec}, we collect the extracted cross sections and average $p_T$ in \mbox{$\sqrt{s} = 7$ TeV} pp collisions obtained using the fits to ATLAS and LHCb data, while the bottom set of results was obtained using CMS and LHCb data.  To obtain these estimates we integrated the fits obtained over $0.1 \leq p_T \leq 80$ GeV including the inherent uncertainty in each state's fit result.  Note that the average $p_T$ for all P-wave polarizations is the same due to the fact that we have assumed that the $\chi_{b0}(mP)$ and $\chi_{b2}(mP)$ states can be obtained from multiplicative scalings of the $\chi_{b1}(mP)$ differential cross section.

As can be seen from Tab.~\ref{tab:sevenxsec}, the extracted $p_T$-integrated cross sections based on ATLAS and CMS data are largely consistent.  We note that the fit to the ATLAS $\Upsilon(1S)$ production data provides a much more constrained prediction than the CMS result due to the fact that the statistical uncertainties achieved by the ATLAS experiment for the S-wave state cross sections were smaller and their results extend to lower $p_T$ than CMS.  Comparing the results obtained for the average $p_T$ using the ATLAS and CMS data, we see that the results are once again consistent with one another within the reported uncertainties, with tension between the two results evident only in the results for the average $p_T$ of $\chi_{b}(1P)$ states.  This stems from the fact that the ATLAS fit decreases more rapidly at low $p_T$ than does the CMS result.  This is reflected in the larger value of the ATLAS fit coefficient $b$ listed in Tab.~\ref{tab:pwavefit}.

\section{Extracted feed-down fractions}
\label{sec:feeddownfracs}

In case of pp collisions, the direct production cross sections can be calculated using the values of the experimentally observed cross sections and feed-down matrix $F$ as follows \cite{Brambilla:2020qwo}
\begin{equation}
   \frac{d^2\vec{\sigma}_{\rm exp}}{dp_T dy} = F \cdot \frac{d^2\vec{\sigma}_{\rm direct}}{dp_T dy} \; ,
   \label{eq:F}
\end{equation} 
where the feed-down matrix $F$ collects the excited bottomonium state branching ratios, and the vectors $\vec{\sigma}_{\rm direct}$ and $\vec{\sigma}_{\rm exp}$ collect the direct and experimentally-observed differential cross sections for the $\Upsilon(1S)$, $\Upsilon(2S)$, $\chi_{b0}(1P)$, $\chi_{b1}(1P)$, $\chi_{b2}(1P)$, $\Upsilon(3S)$, $\chi_{b0}(2P)$, $\chi_{b1}(2P)$, and $\chi_{b2}(2P)$ states, respectively.  The matrix $F$ used herein is listed in App.~\ref{sec:feeddownmatrix}.

Equation \eqref{eq:F} can be applied using any set of $p_T$ bins and integrated on the left and the right.  After this integration, one can invert Eq.~\eqref{eq:F} to obtain the direct production cross sections for all states considered.  Using the direct production cross sections one can then identify the percentage of, e.g., $\Upsilon(1S)$ production in a given $p_T$ bin coming from direct production versus excited state feed down.  For example, at a given $p_T$ and $y$, the feed-down fraction for $\Upsilon(1S)$ production coming from $\Upsilon(2S)$ decays can be computed as 
\begin{equation}
    \left.\left(\frac{d\sigma_{\rm direct}[2S]}{dp_T dy}  \times {\rm Br}[\Upsilon(2S) \rightarrow \Upsilon(1S)\,X]\right)\right/\frac{d\sigma_{\rm exp}[1S]}{dp_T dy} \, ,
\end{equation}
where $X$ represents any additional final particle(s).  If one is interested in the feed-down fractions at a particular value of $p_T$ one can integrate the left and right hand sides in a narrow bin surrounding the value of $p_T$ in question \footnote{Formally one should use a Dirac delta function.}.  Below, we will use this method to extract the feed-down fractions at the average transverse momentum of the $\Upsilon(1S)$ and $\Upsilon(2S)$.

\subsection{Results for feed-down fractions}

In Fig.~\ref{fig:feeddown} we present our final results for the $\Upsilon(1S)$ and $\Upsilon(2S)$ feed-down fractions as a function of $p_T$.  The top row of Fig.~\ref{fig:feeddown} shows our results for the $\Upsilon(1S)$ and the bottom row shows our results for the $\Upsilon(2S)$.  The left column shows results obtained by combining ATLAS S-wave production data and LHCb measurements of the P-wave to S-wave ratios.  The right column shows the result of using CMS data for S-wave production.  Note that, for this Figure, we have summed over the contributions of the $\chi_{b0}(mP)$, $\chi_{b1}(mP)$, and $\chi_{b2}(mP)$ states.  As can be seen from Fig.~\ref{fig:feeddown}, the $\Upsilon(1S)$ feed-down fractions have a significant transverse momentum dependence and, as mentioned in the introduction to this paper, our findings are consistent with those reported previously in Ref.~\cite{Wohri:2015hq}.  The $\Upsilon(2S)$ feed-down fractions only have a strong momentum dependence below $p_T \sim 8$ GeV and, according to our analysis, only weakly depend on the transverse momentum above this scale.

If one evaluates the ATLAS $\Upsilon(1S)$ and $\Upsilon(2S)$ feed-down fractions at the average transverse momentum for the $\Upsilon(1S)$ and $\Upsilon(2S)$ states listed in top set of results in Tab.~\ref{tab:sevenxsec}, one obtains the feed-down fractions listed in the top set of results shown in Tabs.~\ref{tab:avgptfd-1S} and \ref{tab:avgptfd-2S}, respectively.  Evaluating the CMS feed-down fractions at the average transverse momentum for the $\Upsilon(1S)$ and $\Upsilon(2S)$ states listed in bottom set of results in Tab.~\ref{tab:sevenxsec}, one obtains the feed-down fractions listed in the bottom set of results shown in Tabs.~\ref{tab:avgptfd-1S} and \ref{tab:avgptfd-2S}, respectively. For both the $\Upsilon(1S)$ and $\Upsilon(2S)$ our results indicate that, if one uses the average transverse momentum of each of these states, over 75\% of their experimentally observed production cross section comes from direct production.  A consequence of this finding is that the $p_T$- and $y$-integrated suppression of $\Upsilon(1S)$ states seen in the highest energy Pb-Pb collisions, which corresponds to an $R_{AA}$ of approximately 35\% \cite{ALICE:2020wwx,ATLAS:2022exb,Sirunyan:2018nsz,CMS:2023lfu}, cannot be attributed to suppression of bottomonium excited states alone.

\begin{figure}[t]
\begin{center}
\includegraphics[width=0.95\linewidth]{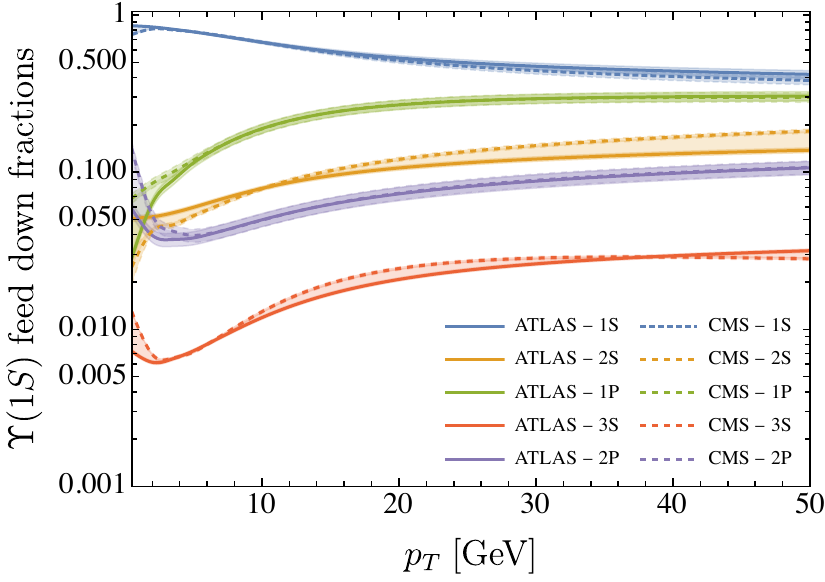}
\\[1em]
\includegraphics[width=0.95\linewidth]{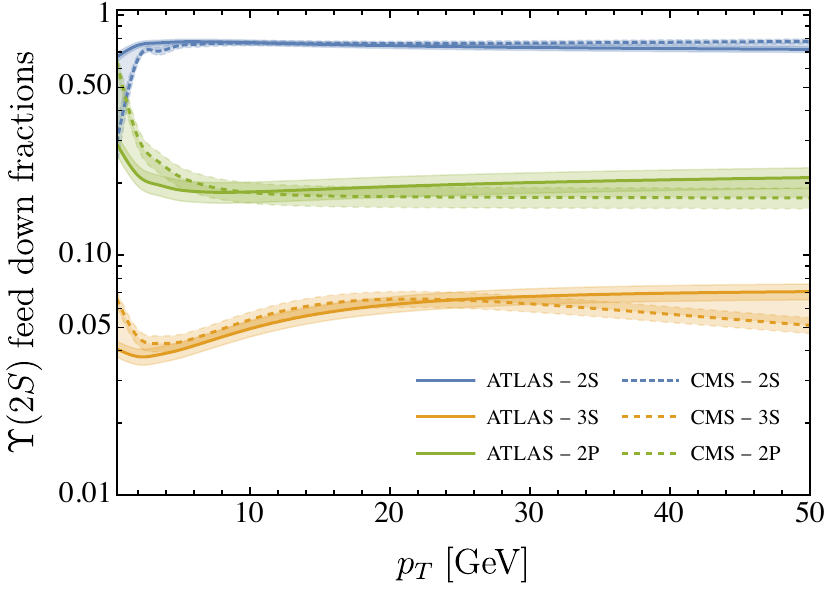}
\end{center}
\vspace{-3mm}
\caption{$\Upsilon(1S)$ (top) and $\Upsilon(2S)$ (bottom) feed-down fractions as a function of $p_T$.  Solid lines indicate results obtained using ATLAS S-wave production data and dashed lines results obtained using CMS S-wave production data. } 
\label{fig:feeddown-comp}
\end{figure}

\begin{table}[t!]
\begin{center}
\begin{tabular}{|c|c|} 
 \hline
 \multicolumn{2}{|c|}{\bf ATLAS + LHCb - 1S} \\
 \hline
 {\bf State} & \; {\bf $\bm{\langle p_T \rangle}$ feed-down fraction} \; \\[0.25em] \hline
$\Upsilon(1S)$  & $0.763\pm 0.010$ \\[0.25em] \hline
$\Upsilon(2S)$  & $0.0625\pm 0.0019$  \\[0.25em] \hline
$\chi_{b}(1P)$  & $0.127\pm 0.009$  \\[0.25em] \hline
$\Upsilon(3S)$  & $0.00786\pm 0.00018$  \\[0.25em] \hline
\;$\chi_{b}(2P)$\;  & $0.039\pm 0.004$ \\[0.25em] \hline
\end{tabular}
\\[1em]
\begin{tabular}{|c|c|} 
 \hline
 \multicolumn{2}{|c|}{\bf CMS + LHCb - 1S} \\
 \hline
 {\bf State} & \; {\bf $\bm{\langle p_T \rangle}$ feed-down fraction} \; \\[0.25em] \hline
$\Upsilon(1S)$  & $0.767\pm 0.010$ \\[0.25em] \hline
$\Upsilon(2S)$  & $0.0561\pm 0.0018$  \\[0.25em] \hline
$\chi_{b}(1P)$  & $0.129\pm 0.009$  \\[0.25em] \hline
$\Upsilon(3S)$  & $0.00778\pm 0.00018$  \\[0.25em] \hline
\;$\chi_{b}(2P)$\; & $0.040\pm 0.004$ \\[0.25em] \hline
\end{tabular}
\caption{$\Upsilon(1S)$ feed-down fractions evaluated at the average transverse momentum of the $\Upsilon(1S)$.  The top set of results was obtained using ATLAS + LHCb data and the bottom set of results was obtained using CMS + LHCb data. }
\label{tab:avgptfd-1S}
\end{center}
\end{table}

\begin{table}[t!]
\begin{center}
\begin{tabular}{|c|c|} 
 \hline
 \multicolumn{2}{|c|}{\bf ATLAS + LHCb - 2S} \\
 \hline
 {\bf State} & \; {\bf $\bm{\langle p_T \rangle}$ feed-down fraction} \; \\[0.25em] \hline
$\Upsilon(2S)$  & $0.774\pm 0.018$  \\[0.25em] \hline
$\Upsilon(3S)$  & $0.0429\pm 0.0032$  \\[0.25em] \hline
\;$\chi_{b}(2P)$\;  & $0.183\pm 0.018$ \\[0.25em] \hline
\end{tabular}
\\[1em]
\begin{tabular}{|c|c|} 
 \hline
 \multicolumn{2}{|c|}{\bf CMS + LHCb - 2S} \\
 \hline
 {\bf State} & \; {\bf $\bm{\langle p_T \rangle}$ feed-down fraction} \; \\[0.25em] \hline
$\Upsilon(2S)$  & $0.758\pm 0.019$  \\[0.25em] \hline
$\Upsilon(3S)$  & $0.0464\pm 0.0035$  \\[0.25em] \hline
\;$\chi_{b}(2P)$\; & $0.195\pm 0.019$ \\[0.25em] \hline
\end{tabular}
\caption{$\Upsilon(2S)$  feed-down fractions evaluated at the average transverse momentum of the $\Upsilon(2S)$.  The top set of results was obtained using ATLAS + LHCb data and the bottom set of results was obtained using CMS + LHCb data. }
\label{tab:avgptfd-2S}
\end{center}
\end{table}

\subsection{Comparison between ATLAS and CMS based results}

In order to assess the systematic uncertainty associated with fitting to either the ATLAS or CMS data, in Fig.~\ref{fig:feeddown-comp} we compare the feed-down fractions obtained using the ATLAS and CMS S-wave production data together with LHCb data for the P-wave to S-wave ratios.  In order to better resolve both large and small feed-down fractions, we present the results obtained on a logarithmic axis, with the ATLAS-based results indicated by solid lines and the CMS-based results indicated by dashed lines.  The darker shaded bands around each result indicate the $1\sigma$ uncertainty while the lighter shaded bands between the ATLAS and CMS results indicate the systematic uncertainty resulting from fits to either the ATLAS or CMS S-wave production data.  As can be seen from this Figure, although there exists some differences between the ATLAS- and CMS-based results, both analyses are largely consistent with one another for $3 \lesssim p_T \lesssim 30$ GeV, which is the transverse momentum range over which nuclear suppression is typically measured. 

\section{Conclusions and outlook}
\label{sec:conclusions}

In this paper we carried out a data-driven extraction of the transverse momentum dependence of bottomonium feed-down fractions using data from the ATLAS, CMS, LHCb Collaborations collected in $\sqrt{s}$ = 7 TeV pp collisions.  We made independent fits to ATLAS and CMS data for S-wave bottomonium differential cross sections, finding both the best fit parameters and quantifying the $1\sigma$ confidence intervals for the fits.  We then made use of LHCb data for ratios of P-wave to S-wave production to construct differential cross sections for the $\chi_{b}(1P)$ and $\chi_{b}(2P)$ states.  Our final results for the fitted differential cross sections were presented in Fig.~\ref{fig:finalxsecs}.  Based on these fits, we computed the $p_T$-integrated S- and P-wave bottomonium production cross sections and average transverse momenta and presented our results in Tab.~\ref{tab:sevenxsec}.

Using these fits and knowledge of the branching ratios for bottomonium excited states to lower lying states, we extracted the feed-down fractions as a function of transverse momentum by inverting Eq.~\eqref{eq:F} to obtain the direct production cross sections as a function of transverse momentum.  This then allowed us to compute the percentage of the experimentally-observed differential cross sections for the $\Upsilon(1S)$ and $\Upsilon(2S)$ states coming from feed down of higher-lying excited states.  At all stages of this process, we propagated uncertainties associated with the fits obtained and all relevant branching ratios.  Our final results for the transverse momentum dependence of the $\Upsilon(1S)$ and $\Upsilon(2S)$ feed-down fractions using either ATLAS or CMS S-wave production data were presented in Fig.~\ref{fig:feeddown} and compared to one another in Fig.~\ref{fig:feeddown-comp}.

We found that there is significant $p_T$ dependence of the $\Upsilon(1S)$ feed-down fractions, while the $\Upsilon(2S)$ feed-down fractions became approximately independent of transverse momentum for $p_T \gtrsim$ 8 GeV.  Based on these extractions, we then computed the feed-down fractions at the average transverse momenta for the $\Upsilon(1S)$ and $\Upsilon(2S)$ states presented in Tab.~\ref{tab:sevenxsec}. We found that, when evaluated at the average transverse momentum, direct production of both the $\Upsilon(1S)$ and $\Upsilon(2S)$ was approximately 75\%.  This indicates that the suppression of the $\Upsilon(1S)$ seen in heavy-ion collisions cannot be solely explained by suppression of bottomonium excited states since, when integrated over $p_T$ and $y$, nuclear suppression of the $\Upsilon(1S)$ is approximately 35\%.  In order to allow other research groups to make use of our analysis, we have provided tab-delimited data files corresponding to the results shown in Fig.~\ref{fig:feeddown}.  These are included as ancillary/supplemental files.

We note that, although the data analyzed came from $\sqrt{s} = $ 7 TeV collisions, experimental measurements of bottomonium production cross sections are found to scale approximately linearly with $\sqrt{s}$.  As a result, one can estimate the cross sections at $\sqrt{s}$ = 5.02 TeV by scaling the cross sections by $5.02/7 \sim 0.717$. The resulting 5.02 TeV cross sections are compatible with previous estimates, see e.g. \cite{Brambilla:2021wkt}. Additionally, if one assumes linear scaling in this range of collision energies, then the feed-down fractions provided by our analysis can also be used in phenomenological analyses of $\sqrt{s}$ = 5.02 TeV collisions.

Looking forward, one can immediately apply the feed-down fractions extracted here to phenomenological computations of bottomonium suppression in heavy-ion collisions. In terms of improving the analysis presented here, it would be ideal to have additional experimental data on P-wave production, in particular the $\chi_{b0}(mP)$ states, however, this would be challenging due to the small branching ratio of this state to $\Upsilon(nS)\gamma$.  Finally, in order to reduce the uncertainty associated with the extrapolation from 7 to 5.02 TeV, it would be ideal if experimental measurements of bottomonium production could be performed at this collision energy or a collision energy closer to this value.

\vfill\null 

\acknowledgments{
We thank H.S. Chung for discussions.  This work was supported by the U.S. Department of Energy, Office of Science, Office of Nuclear Physics through the Topical Collaboration in Nuclear Theory on Heavy-Flavor Theory (HEFTY) for QCD Matter under award No.~DE-SC0023547 and by U.S. Department of Energy award No.~DE-SC0013470.
}


\appendix

\begin{widetext}

\section{Feed-down matrix components}
\label{sec:feeddownmatrix}

Here we list the components of the matrix $F$ which appears in Eq.~\eqref{eq:F}.  The values in this matrix can be obtained using the branching ratios for bottomonium states listed in the PDG \cite{pdg}.
\begin{equation*}
{\fontsize{7}{10}\selectfont
F = \begin{pmatrix}
1 & 0.265\pm 0.005 & 0.0194\pm 0.0027 & 0.352\pm 0.020 & 0.180\pm 0.010 & 0.0657\pm 0.0015 & 0.0038\pm 0.0017 &
   0.115\pm 0.011 & 0.077\pm 0.009 \\
 0 & 1 & 0 & 0 & 0 & 0.106\pm 0.008 & 0.0138\pm 0.0030 & 0.181\pm 0.019 & 0.089\pm 0.012 \\
 0 & 0 & 1 & 0 & 0 & 0 & 0 & 0 & 0 \\
 0 & 0 & 0 & 1 & 0 & 0 & 0 & 0.0091\pm 0.0013 & 0 \\
 0 & 0 & 0 & 0 & 1 & 0 & 0 & 0 & 0.0051\pm 0.0009 \\
 0 & 0 & 0 & 0 & 0 & 1 & 0 & 0 & 0 \\
 0 & 0 & 0 & 0 & 0 & 0 & 1 & 0 & 0 \\
 0 & 0 & 0 & 0 & 0 & 0 & 0 & 1 & 0 \\
 0 & 0 & 0 & 0 & 0 & 0 & 0 & 0 & 1 \\
\end{pmatrix} .
}
\end{equation*}
The components of this matrix are, for $i < j$, given by $F_{ij} = {\rm Br}(j \rightarrow i X)$, where $X$ represents any additional final state particle(s) and $i$ and $j$ represent states in the list $\Upsilon(1S)$, $\Upsilon(2S)$, $\chi_{b0}(1P)$, $\chi_{b1}(1P)$, $\chi_{b2}(1P)$, $\Upsilon(3S)$, $\chi_{b0}(2P)$, $\chi_{b1}(2P)$, and $\chi_{b2}(2P)$.  For $i > j$ the components are zero since these components correspond to vacuum transitions from lower-lying states to higher-lying states.  The diagonals are unity since to obtain the experimentally observed cross section, one divides the observed yield for a given state by its corresponding dimuon branching fraction, making these unity by definition.
\end{widetext}

\bibliographystyle{apsrev4-1}
\bibliography{main}

\end{document}